# Demonetization and Its Impact on Employment in India

Dr. Pawan Kumar*

Assistant Professor, Ramjas College,
University of Delhi, Delhi-110007, India

On November 08, the sudden announcement to de-monetization the high denomination currency notes of Rs. 1000 and Rs. 500 sent tremors all across the country. Given the timing, and socio-economic and political repercussions of the decision, many termed it a ‘financial emergency’. Given high proportion of these notes in circulation (over 86 percent), ‘demonetization’ led to most economic activities, particularly employment, affected in a big way. Political parties, however, seemed divided on the issue, i.e. those in favor of the decision feel it will help to curb the galloping size of black money, fake currency, cross boarder terrorism, etc. In sharp contrast, the others believe it is a purely un-thoughtful, decision, based on no or poor understanding of ‘black economy’, and hence is only politically motivated in wake of the assembly elections due in a couple of states particularly UP. In addition, they also believe, the given decision of ‘demonetization’ is far from ground reality, and hence is unlikely to fetch anything tangible except causing miseries to millions of people by making them stand in queue day in and day out.

As mentioned above, among all implications of ‘demonetization’, its impact on employment is important, particularly in a situation when majority of wage payments are made in cash form. Given the nature of huge informal employment, more than 95 percent of total transactions in India are in cash form (Live Mint, January, 01, 2017). The decision of sudden ‘demonetization’ therefore led the labor market dynamics changed significantly by rendering millions of workers exposed to increased uncertainty in employment; they resorted to ‘reverse migration’. For employment growth, economic growth is one of the important factors. In a situation, when the recession phase caused by global crisis of 2009 had hardly dimmed away, the recent move of ‘demonetization’ will push growth downward as predicted by most economic agencies including RBI and IMF. The prediction of decline in GDP ranges from 0.5 percent to 2 percent. Theoretically, a unit decline in growth rate results a decline in employment growth rate, a concept called as employment elasticity (EE) of output. So, labour market, particularly informal in nature, will be most affected by the move.

*pawankumar@ramjas.du.ac.in

Before, evaluating the employment impact of the recent move of ‘demonetization’, it is imperative to summarize the labour market in India. Following the Report on Fifth Annual Employment and Unemployment Survey (2015-16), a couple of points are worth mentioning. *First*, among all, very few households (20 percent) with monthly income less than or equal to Rs. 10,000 have bank account. Seconds, majority of workers belong to this income group only; 82 percent among self-employed, 60 percent among regular or salaried workers, 87 percent among contract workers, and 96 percent among casual workers. *Third*, majority of workers, in all category of employments have no written contract, 65 percent (regular workers), 68 percent (contract workers), 95 percent (casual workers). *Fourth,* majority of workers (71.2 percent) receive no social benefits. *Fifth,* 60 percent of workforce belongs to just six states, Tamil Nadu, Maharashtra, Andhra Pradesh, Gujarat, Karnataka and Uttar Pradesh.

For quite sometimes, the labour market in India has been witnessing numerous uncertainties including the problem of world recession, and growing ‘automation’ particularly in the manufacturing sector. More precisely, in the last one and a half decade, India has emerged a global power in terms of the development or diffusion of new technology in the form of ICT. ICT intensity, defined as the ratio of ICT investment to non ICT investment, has increased significantly across industries led to ‘automation’ in most production (and distribution). Its impact on productivity led growth, and direct employment is well documented. However, its negative employment impact, particularly in the ICT using manufacturing sector has largely been ignored. So, in a situation, when the debate, whether the net employment impact of ICT on the economy as a whole is positive, is still un-conclusive; any major policy change like ‘demonetization’ is likely to make the employment scenario further volatile by causing uncertainties to rise in labour market, mainly the informal employment.

Informal employment, which constitute as high as 95 percent of all employment is backed with no (or least) social security such as health, education or provident fund benefits. Workers are subject to be fired (or lay-off) at any point of time during the production (or distribution) process. Since majority of wage payment is made in cash form; they are thus the ones to face misery caused by the recent announcement of ‘demonetization’. According to

*pawankumar@ramjas.du.ac.in

Financial Express (Nov, 24, 2016), an estimated 4 lakh workers, largely belonging to this segment will be affected by the decision.

Further, a point worth mentioning is that, all employment in the formal sector is not formal per se; rather informal in nature, i.e. not backed by social security. It rose from 38.7 percent to 46.6 percent during 2000 to 2005, or a phenomenon known as '*infomalization of the formal sector*' (NCEUS, 2009). Or, social security as percentage of the value of output in the formal sector has recorded a perceptible declined, from 14 percent in 2000 to 10 percent in 2013 (ASI, 2013-14). So, if output growth declines by the recent policy of de-monetization, as predicted by most economic agencies including World Bank, employment of largely these informal workers will be affected. Further, using the same data source, in the last ten years since 2003, a total of 94163 additional factories gave rise to nearly 4 million employment generations; or 41 workers per factory. So, it is easy to infer, if de-monetization' led a factory shuts down; it will render nearly 41 people unemployed. Further, from the sectoral analysis of industries within the manufacturing sector a couple of points also are worth mentioning.

It is further found that labour intensive units such as food and beverage, tobacco, textile, leather, wood and jewelry employ nearly half of the total workers in the organized manufacturing sector of the economy (ASI, 2013-14). Given that nearly 84 percent of total factories have employment in the range of 0 to 99 are thus prone to be affected by the recent move of the government. Newspaper, electronic media or social media are flooded with the news on 'reverse migration', i.e. lakhs of people are forced to flee the industrialized state such as Punjab, Haryana, Maharashtra, Gujarat etc., to their place of origin.

Since in some of these industries, share of female workers is higher, wearing apparel, except fur apparel (46 percent), knitted and crocheted apparel' (36.08 per cent) and other food products' (33.98 per cent). Since, women workers are inherently weak in bargaining power and hence are subject to more labour market uncertainties in terms wage rate or social security. In these industries, as mentioned above, cash payment is generally the only mode of payment; they are thus will be affected more than their male counterpart by any policy change such as 'demonetization'.

Further, using the principle of employment elasticity (EE), an attempt is made to estimate likely impact of 'demonetization on employment in the organized manufacturing sector of India.

*pawankumar@ramjas.du.ac.in

Over the past half a decade, owing to numerous reasons including ICT led automation; employment elasticity has declined perceptibly in most industries. Employment elasticity, by definition, measures percentage change of labour demand (or employment) due to a percentage change in output level. Numerically, an EE equal to one (say) means – an output growth rate of 10 percent (say) resulted in a 10 percent growth in employment, and vice versa. In short, it measures the employment intensity of a unit of output produced. In the past, some sectors like construction or ICT witnessed unit level of EE. The same principle of EE is used to measure the extent of employment loss due to an expected demonetization led decline in output growth rate.

According to Economic Survey (2013-14), for the country as a whole, EE rose marginally from 0.16 during 1994-00 to 0.19 during 2000-12 for all sectors, i.e. including both formal and informal. Using the ASI data, it is found that during 2010-2014, 21 industries in organized manufacturing sector, 14 percent output growth rate and 3.2 percent rate of employment growth gave a 0.21 level of EE. Two scenarios are presented, first assuming an extent of decline of 30 percent lower than expected output growth, and the second with a decline of 20 percent lesser than expected output in the next year 2017-18. *Assuming the same growth rates of output and employment respectively (or the same level of EE) for next three years (2015-2017), scenario one will render nearly 1 million workers unemployed; this is expected to be nearly 6.4 Lakhs in 2017-18 in these industries alone. It is also found that the employment impact of de-monetization is not uniform across industries. More precisely, labour intensive industries such as food, textile, wearing apparel, leather and leather product industries are seemed to have witnessed the major loss, i.e. these industries will register an absolute loss of employment with 32502, 42072, 37183 and 38089 respectively (in scenario one) or, 21667, 28048, 24788 and 25399 respectively (in scenario two)*

Now coming to the informal sector, according to a Report by ASI (2010-11), roughly a fifth of the almost 32 million people employed in the textile and garment sector, are daily wage earners. Hence, any policy change impacting decline in output growth makes these people be affected more. Further, according to NCEUS, 2009 Report, since majority of people (78.7 percent) belonging to informal sector are poor, or constituting 90 percent of casual workers and 75 percent of self-employed people. So, these are the ones who bear the major burnt of the decision of 'demonetization'.

*pawankumar@ramjas.du.ac.in

A fact, well-documented, is that the formal sector has reached its saturation point of employment; it is thus unable to help in additional employment generation. Qualitatively also, things are not all rosy here, i.e. the percentage of income spent on social security has steeply gone down over the years. A tendency of quite a high proportion of contractual workers is found prevalent in most sectors, for instance, remediation activities and other waste management services (100%), Waste collection (91.70 %) and Mining and quarrying (83.89 per cent) and so on (ASI, 200-10). At factory level, 26.42 per cent factories were reported to have employed contract workers in 2009-10; found to be highest in public sector (35.02 per cent), 39.56 per cent in Joint Sector and 26.18 per cent in Private Sector (ASI, 2013-14). At the State level, highest is found to be in in Tripura (67 percent), 58.50 per cent in Bihar, 47.19 per cent factories in Nagaland and 45.06 per cent factories in Dadra & Nagar Haveli. Under Public Sector, the highest is found in Chandigarh (around 80 percent) followed by Chhattisgarh (70 percent) and Rajasthan (56 percent).

From the above analysis, it can be easily concluded that employment scenario in the country is not conducive enough to face any challenge such as the ‘demonetization’ of currency. In a country, when 79 percent of non-agricultural wage workers have no written contract and only one fourth are eligible for any social security, the decision is certainly a cause of concern. India has the world largest youth population, so for any developing country like India, it is the time to harness the population dividend by providing them gainful employment. No doubt, impact of ICT on growth and direct employment is well documented, but its indirect negative employment impact ICT using manufacturing sectors can-not be ignored. Given this, the decision of ‘demonetization’ will further destabilize the already volatile labour market in India.

*pawankumar@ramjas.du.ac.in